# Spin-polarized Voltages on a 2D Self-assembled Plasmonic Crystal

Nicholas V Proscia[1,2,3], Matthew Moocarme[1,2], Roger Chang[4], Ilona Kretzschmar[4],
Vinod M. Menon[1,3] & Luat T. Vuong[1,2]

[1]Dept. of Physics, Graduate Center of the City University of New York (CUNY), United States
[2]Dept. of Physics, Queens College of the City University of New York (CUNY), United States
[3] Dept. of Physics, City College of New York of the City University of New York (CUNY), United States
[4] Dept. of Chemical Engineering, City College of New York of the City University of New York (CUNY), United States
Email: luat.vuong@qc.cuny.edu

**Abstract:** The Photon Drag Effect (PDE) is a nonlinear process akin to optical rectification in which the momentum of light is transferred to charged carriers and converted to a DC voltage. Here, we experimentally demonstrate the spin-polarized voltage—associated with the transference of light's spin angular momenta to the linear momenta of charges— with visible-light illumination on a nanovoid self-assembled plasmonic crystal surface. Numerical calculations show that the gradient force, generally considered independent of polarization, is responsible for the majority of the momentum transfer. The PDE in this achiral system represents a distinct spin-orbit interaction that produces asymmetric hotspots whose locations change with circular polarization handedness. Our results significantly advance our understanding of the PDE and demonstrate realistic potential for scalable plasmonic materials that utilize PDE.

## 1. Introduction

The momentum transfer from light to matter is a well-known phenomenon that is documented as early as the turn of the 20th century.[1] A modern variant of this phenomenon is the Photon Drag Effect (PDE), which describes the time-averaged force exerted by photons onto charged particles.[2-4] The PDE is a nonlinear second order intensity-dependent phenomenon[2,4] that generates a time averaged DC voltage across the material. The PDE voltages that result from the illumination of electromagnetic fields on smooth conductor or semiconductor surfaces are extremely small *i.e.,* limited in its value for practical applications.[2,4,5] In the presence of surface plasmon polaritons (SPPs), however, the corresponding photo-induced voltages are enhanced by several orders of magnitude.[6] The subject of photo-induced potentials associated with the PDE in nanostructured plasmonic metals or "plasmo-electric" response has subsequently gained significant attention.[7-16] Although, the direction of the PDE voltage is generally in a direction parallel to the plane of incidence,[11,12] as expected from conservation of linear momentum principles,[2,3] there can be an appreciable voltage produced in a direction *perpendicular* to the incident plane, known as a transverse photo-induced voltage (TPIV).[11,12,17]

The TPIV results from the conversion of spin angular momenta (SAM) of circularly-polarized light (CPL) to the linear momenta of charges, whose direction changes with spin or CPL handedness.[11,12,17] An explanation of the TPIV is associated with the longitudinal wavevector that shifts with geometric phase and provides the kernel for generalized electromagnetic spin-orbit interactions.[18] With plasmonic materials, the spin-dependent wavevector of the CPL couples to "spin-locked" SPPs, whose henceforth preferential transversal momentum is dictated by spin.[19,20] A second, more classical explanation ascribes the TPIV to a polarization-dependent Lorentz force that occurs via scattering[13]. This plasmo-electric model that incorporates the Lorentz force provides good agreement with experimental trends. While the latter provides a framework for addressing opto-mechanical forces,[21,22] the former, which emphasizes the topological features in the longitudinal component of the electric fields, connects with nonplasmonic phenomena.[23-25] In our investigation of the TPIV, we provide a dual perspective of the geometric phase in Lorentz forces with spin-orbit interactions in an analysis that bridges the two frames above.

Here, we numerically and experimentally demonstrate the PDE TPIV in a bottom-up fabricated 2D plasmonic crystal composed of nanovoids. We find that the spin-polarized voltage is largely attributed to the presence of antisymmetric hotspots in each lattice whose locations change with incident CPL handedness. We analytically represent the superposition of fields as a distinct class of spin-orbit interactions, where the incident SAM couples to its extrinsic orbital angular momenta (EOAM)[26] that results from the contoured surface of nanovoid film. The spin-polarized voltages are enhanced when the surface plasmon polariton (SPP) resonance approaches the plasma frequency of the material and maximal PDE voltages are produced at a wavelength that is red-shifted from the absorption peak, in agreement with numerical simulations. Our investigation represents the first study of the plasmo-electric response in a bottom-up fabricated plasmonic crystal.

## 2. Numerical Model

Our numerical model for the PDE is associated with the Lorentz force, which governs the motion of the conduction electrons through the near-field interactions with the incident electromagnetic wave[3]. Its time-averaged relation to the complex amplitude of the electric field $\boldsymbol{E}$ is[7,11,13]

$$\mathbf{F} = \frac{\alpha_R}{4} \nabla |\boldsymbol{E}|^2 + \frac{\alpha_I}{2} \text{Im}\{\sum E_j^* \nabla E_j\}, \qquad (1)$$

where $\alpha_R$ and $\alpha_I$ are the frequency-dependent real and imaginary parts of the material polarizability, respectively. In Eq. 1, the first term is referred to as the gradient force (GF) while the second term is known as the scattering force (SF).[13] These two terms govern the dynamics of the PDE. The SF depends strongly on polarization; the GF is generally considered to be polarization independent as it is dependent on the absolute value of $\boldsymbol{E}$.

In fact, the GF carries a critical role in generating the TPIV even though the TPIV depends strongly on the polarization of $\boldsymbol{E}$: the intensity of $\boldsymbol{E}$ in a direction normal to the surface is a superposition of incident and scattered fields that depends on the topological phase associated with circular polarization handedness. The SF is proportional to the imaginary part of the product of the electric-field with its gradient and represents the difference in phase between two orthogonal field components. The GF is the dominant term in generating the TPIV; the gradient of $|\boldsymbol{E}|^2$ is typically much larger than $Im\{E_j^* \nabla E_j\}$.

We numerically calculate the PDE voltage on our plasmonic crystal surface by utilizing the dipole approximation model.[13] The voltages on our nanovoid structure are numerically simulated with Comsol Multiphysics finite element analysis software. The simulation domain is a rhombic unit cell representing a 2D hexagonal lattice. The angle of incidence (AOI) is in the *x-z* plane and range from 21° to 51°.

We incorporate the PDE as a perturbation and calculate the voltages produced across the plasmonic nanovoid sample from the local forces on individual structures. This is a reasonable approximation as intensity-dependent changes in refractive index are small.[27] Adjacent nanovoids in the *y*-direction (*x*-direction) are treated as an assemblage of batteries in series (parallel).[9] Therefore, the cumulative voltage is product of the average voltage across a single unit cell with the number of unit cells in the transverse direction of the illuminated area.

$$TPIV = \frac{1}{e} * \frac{1}{Vol} \int F_y(\boldsymbol{r})\, d\boldsymbol{r} * L \qquad (2)$$

where *e* is the charge of an electron, *Vol* is the volume of the unit cell used in the simulation, *L* is the length of the incident beam in the *y*-direction. The combination of Eqs. (1) and (2) yield the final numerically-calculated TPIV due to CPL incident on the nanovoid film.

Here we measure that the TPIV generated from the GF is significantly larger than that generated from the SF. The high field localization associated with the presence of plasmons leads to large TPIV via the GF and PDE is described in Secs. 4 and 5.

## 3. Experimental Procedures

The self-assembled nanovoid plasmonic crystal is a textured gold film with truncated spherical voids in a 2D hexagonal lattice.[28] The surface is fabricated as follows [Fig. 1(a)]: An initial gold film is deposited via physical vapor deposition on a glass substrate with a nominal Ti wetting layer for adherence. A monolayer of close-packed 600 nm diameter polystyrene (PS) spheres is self-assembled on the gold film via the process of convective assembly.[29] The film is then placed in a gold-plating bath where gold is electro-deposited between the PS spheres and forms a gold cup around each PS sphere. Finally, the PS spheres are removed by dissolution in tetrahydrofuran, which then leaves the periodic array of truncated spherical nanovoids. The height of the gold deposit around each sphere is controlled by the total amount of charge that passes through the bath. The depths of the nanovoids are moderately homogeneous and ranged from 75 nm to 105 nm across the sample. The median nanovoid depth of 90 nm is used for numerical calculations of the TPIV. The irregularity in the gold deposition depth is considered to be the primary source of sample inhomogeneity in our experiments.

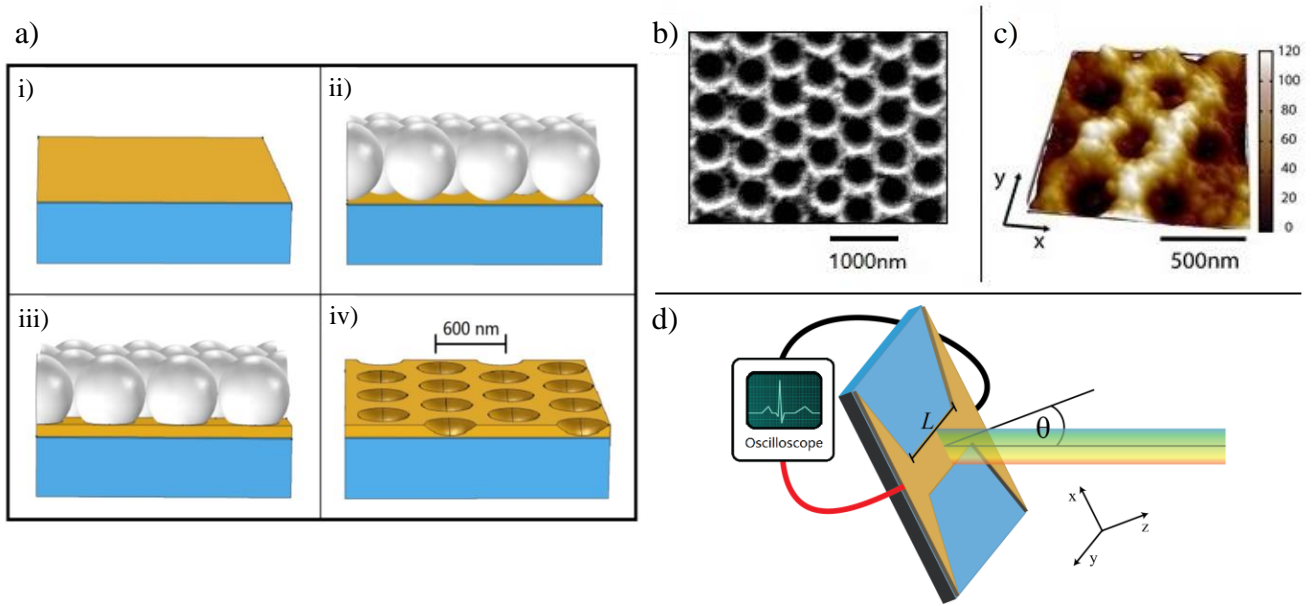

Fig. 1.(a) Principal stages of the fabrication process for the nanovoid surface: (i) 20 nm layer of gold. (ii) PS spheres self-assembled on gold film. (iii) Gold is electroplated onto the gold film underneath. (iv) The PS spheres are removed. (b) SEM and (c) AFM image of nanovoids. The nanovoids had median depth of 90 nm, a corresponding rim diameter of 430nm, and a lattice constant of 600 nm. (d) Schematic of the sample setup, where $\theta$ is the AOI in X-Z plane. The sample area is 3 mm in the x-direction, and 14 mm the y-direction, transverse to the plane of incidence. The voltage is measured across the y–axis of the sample.

The nanovoid sample is illuminated with a tunable optical parametric oscillator that is pumped with a frequency-tripled Q-switched Nd:YAG laser [Fig. 1(d)]. The pulses are 4.5 ns in duration and have a repetition rate of 10 Hz. The DC voltages are detected and measured with an oscilloscope. The sample is electromagnetically shielded from the ambient environment. The active sample area has a cross section of 3 mm in the x-direction and a width of 14 mm between the two electrodes in the y-direction.

The generated TPIV is independent of the width of the beam in the x-direction and proportional to the length of the beam in y-direction. In order to measure the largest voltage with minimal damage to the samples, a highly elliptical beam profile is used. All voltage measurements and calculations in this experiment are normalized to an intensity $I = 1\ MW/cm^2$ and a length $L= 7$ mm in the y-direction. The area of the beam spot size on the sample is also taken into account in order to consider the effective intensity associated with the various measurement angles.

## 4. Results and Discussion

Figure 2 shows experimentally-measured reflectivity data for CPL illuminating the nanovoid sample between 490 nm and 590 nm. The numerically-calculated dispersion curves are overlaid. The theoretical TM and TE dispersion curves for the nanovoid sample have been investigated in Ref. 30. The spectral properties of this mode are independent of the incident CPL polarization handedness. We center our investigation on the plasmon mode that occurs between 500 nm and 600 nm for CPL.

The depth of the individual nanovoid affects the spectral response of the SPP mode: as the nanovoid depth increases, the SPP mode blue-shifts [Fig. 2]. Variations in nanovoid depth in our samples are evident in the broadening of our measured plasmonic absorption, which increases at higher AOIs and leads to a less-pronounced absorption mode at higher AOI. We attribute the unusual red-shift of the SPP mode to incomplete removal of the PS spheres.

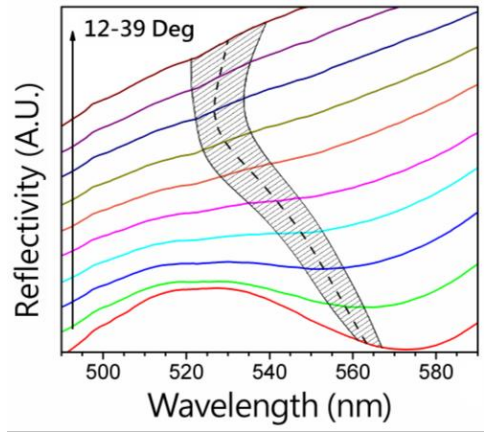

Fig. 2. Experimentally-measured reflection spectra of CPL from nanovoid sample. The dashed line and shaded area show the SPP dispersion from numerical simulations corresponding to a nanovoid depth of 90 ± 15 nm.

The experimental TPIV is compared with numerical calculations in Fig. 3. Right circular polarizations (RCP) and left circular polarizations (LCP) produce voltages of opposite signs while TM-polarized light produces a significantly reduced TPIV [Fig. 3(a)]. The incongruous doubling of the measured TPIV for LCP could be attributed to the additional EOAM associated with the highly elliptical spatial profile.

An interesting feature of the RCP TPIV is an enhancement of the voltage at increasing angles [Fig. 3(b) & 3(c)]. The TPIV increases linearly for the AOIs below 33°, and then near 33°, the TPIV saturates and with only a slight increase at AOIs above 33°. We believe that the enhancement and subsequent saturation of the TPIV lies in the overlapped spectral response of this SPP mode with the plasma frequency of gold, or a wavelength of approximately 500 nm. In the linear regime, *i.e.* an AOI for 21° to 33° [Figs. 3(c-d)], the SPP mode correspondingly shifts towards higher energies [Fig. 2], and therefore closer to the plasma frequency of the gold. At AOIs higher than 33°, the small dispersion of the plasmon mode leads to a nearly constant TPIV and coincides with constant spectral separation between the plasmon mode and the plasma frequency. We are not the first to consider that the TPIV may increase as the SPP mode approaches the plasma frequency; a higher degree of momentum transfer between the SPPs and conduction electrons are first reported to occur via SPP-electron scattering in Ref. 13. Our observations above support the conjecture that relates the magnitude of the TPIV to the overlap between the SPP's spectral response and bulk plasma frequency.

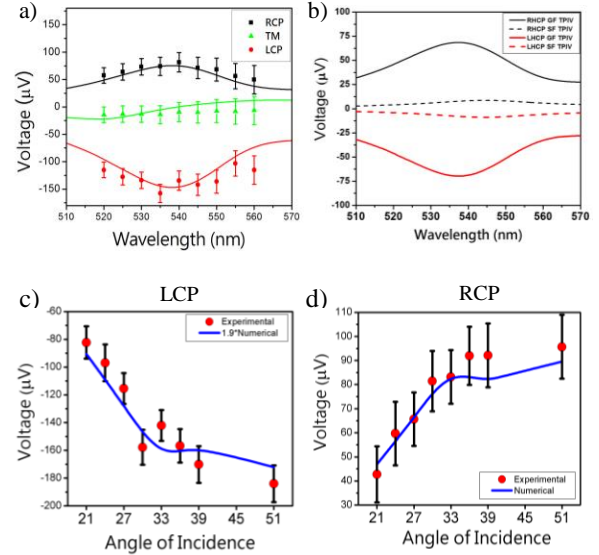

Fig. 3. (a) Experimental (data points) and numerically calculated (solid lines) showing the TPIV at an AOI of 30°. The Numerical LCP voltages are scaled by a factor of 1.9 (b) The GF (solid line) and SF (dashed line) contributions to the TPIV for both LCP and RCP at an AOI of 30°. (c) LCP and (d) RCP peak TPIV as a function of angle. The red dots are experimental data while the blue line represents the peak TPIV from numerical calculations.

What is surprising is that the GF is an order of magnitude larger than the SF and dominates the contribution of the Lorentz Force in our numerical simulations [Fig. 3(b)]. The large net GF is attributed to the asymmetric hot spot that develops on each nanovoid when illuminated with CPL. The hotspot shifts in location with AOI and when illuminated with the opposite-handed CPL and its asymmetric position is the primary driver for the polarity of the TPIV. We illustrate the electric-field amplitude on the surface of the nanovoid structure [Fig. 4], which shows the asymmetric hotspot.

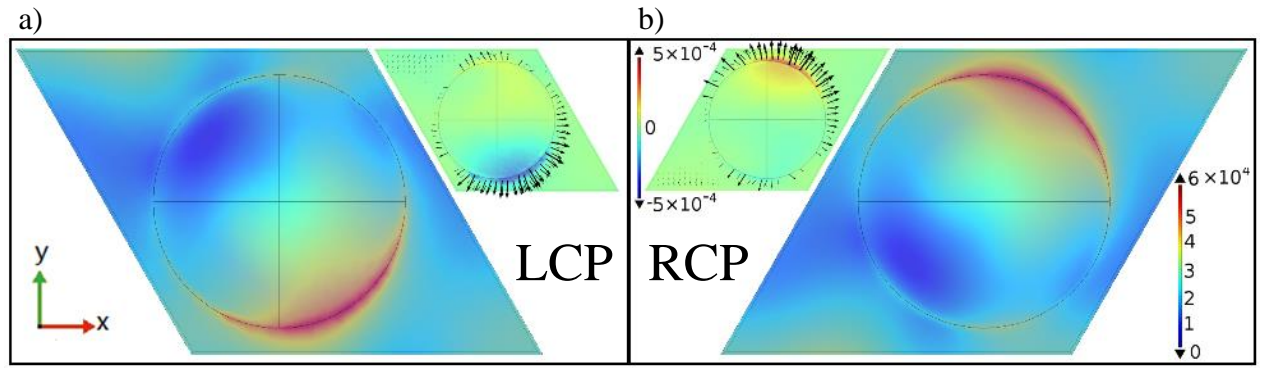

Fig. 4. Contour plot of the E-field normal on the surface of the nanovoid structure when illuminated with (a) LCP and (b) RCP at the plasmon resonance for an AOI of 30°. The black arrows are the E-field projected on the surface of the intervoid plane. The insets are a contour plot of the transverse force along the surface of a nanovoid unit cell for LCP (left) and RCP (right). The arrows indicate the net force at the tail of arrow.

Our observation of the polarization-dependent movement of hotspots on an *achiral* structure is novel and differs distinctly from prior observations that either identify polarization-dependent hotspot locations when light illuminates an anisotropic plasmonic structure or when light is imprinted with optical vortices or intrinsic orbital angular momentum (IOAM).[26] In prior work, the asymmetric hot spots are viewed as a manifestation of coupled TE and TM modes[16] or spin-orbit interactions.[18,26,31,32] Here, we classify the polarization-dependent chiral intensity patterns from an achiral plasmonic structure in the framework of spin-orbit interactions.[11]

The direction of the GF on the nanovoid structure is static in a manner that the net voltage due to the GF does change when integrated over the entire unit cell. The magnitude of the force is greatest at the location of the local void hotspot. The inset plots of Fig. 4 show colormaps and vectors of the magnitude and direction of the time-averaged force experienced by the electrons in the plane of the plasmonic surface. The majority of the time-averaged force is created at the rim of the nanovoid and is directed in the radial direction away from the rim edge. The radial forces indicate that appreciable magnetic fields are produced perpendicular to the nanovoid surface.[27]

## 5. Analysis

A simple analysis illustrates the underlying GF of the TPIV with connections to spin-orbit interactions *i.e.*, changes in the longitudinal scattered field component that depend on the CPL handedness. From Ref. 31, the longitudinal field $\Delta_\pm$ of a ±-handed CPL beam with a transverse profile of $A' = A(\rho, \phi, z')$ is:

$$\Delta_\pm = -\int_{-\infty}^{z} e^{-ik(z'-z)} \left[\partial_\rho A' \pm \left(\frac{i\partial_\phi}{\rho} A'\right)\right] dz' e^{\pm i\phi}. \quad (3)$$

Prior examples of spin-orbit interactions with plasmonic structures address the real part of the square-bracketed term, *e.g.*, the combination of CPL scattering from round apertures imprinted with IOAM [31-33]. Here we adjust the model of Ref. 31 and consider the oblique plane-wave illumination of a round aperture. The hotspots are the intensity patterns of the electric fields normal to the plasmonic-surface, which contain contributions of both transverse and longitudinal electric fields:

$$E_p = -\cos(\theta)\Delta_\pm + \sin(\theta) A \quad (4)$$

where the unit-normalized transverse profile $A$ is a Heaviside function in the oblique plane defined by $\rho\cos(\phi) = z\tan(\theta)$

$$A = H(\rho - \frac{a}{\sqrt{(1+\cos^2(\phi)\tan^2(\theta))}}), \quad (5)$$

where $a$ is the radius of the aperture. In our system, the spin-orbit interactions arise from the imaginary component of the square-bracketed term in Eq. 3, which now represents the coupling between azimuthal changes in intensity or EOAM, $Im\left[\left(\frac{i\partial_\phi}{\rho}A\right)\right]$, and radial variations phase, $Im[\partial_\rho A]$.

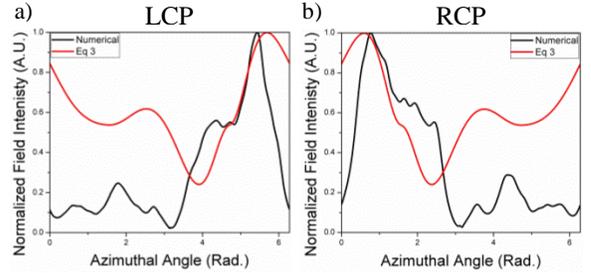

Fig. 5. The normalized z-component of field intensity along the rim of a circular aperture according to Eq. 3 (red) and calculated from the numerical simulation along the rim of the nanovoid(black) for (a)LCP and (b)RCP. The azimuthal angle represents the angular position around rim of each structure.

In Fig. 5, we plot $|E_p|^2$ [Eq. 4] around the rim of the metallic circular aperture at an AOI of 30° and ratio of aperture radius to incident wavelength of 215/543 (red). The numerically-simulated intensity around the nanovoid for the same AOI is also shown (black). The field intensity perpendicular exhibits mirror symmetry along the plane of incidence. Differences in the simulation mesh for LCP and RCP lead to variations from perfect mirror symmetry for the numerical results. The simple model [Eq. 3] assumes that the profile $A(\rho, \phi, z)$ is produced by a perfect absorber and neglects the contribution of reradiated fields. Nonetheless, a comparison of analytical and numerical results indicates agreement in the azimuthal location of the hotspots and change in location between RCP and LCP. This simple model may be generalized to understand spin-orbit interactions that occur via scattering in other subwavelength geometries.

## 6. Conclusion

In summary, we experimentally and numerically demonstrate the TPIV created in self-assembled nanovoid plasmonic gold film when illuminated with CPL in the visible spectrum. The transfer of momentum is expressed through the GF and due to a symmetry breaking on the plasmonic sample when CPL illuminates at non-normal angles of incidence.

The TPIV is largely associated with GF and asymmetric hotspots that change in location with polarization handedness. The polarization-dependent hotspots represent a class of electromagnetic spin-orbit interactions where the radial changes in phase are coupled to the azimuthal changes in intensity. The rim of the nanovoid at oblique AOI produces the break in radial and azimuthal symmetry in the electromagnetic fields, which lead to asymmetric hotspots and the TPIV associated with GF.

Further control or enhancement of the PDE may lead to the design of novel detectors that achieve distinctive advantages over common semiconductor technology, for several reasons. Firstly, the response time of the PDE voltage is extremely fast—only limited by the decay time of the SPP, which is on the order of tens of femtoseconds.[6] Moreover, the photo-induced voltages depend on the plasmonic response of a material, whose spectral absorption and dispersion can be designed and tailored. The measurable differences associated light spin angular momentum (SAM) that are demonstrated here via the polarization-selective dynamics of the PDE may provide a path towards the realization of spin-polarized detectors.[11,16,17] Lastly, the scalability of the bottom-up fabrication of the nanovoid 2D hexagonal plasmonic crystal employed here could be a candidate for future commercial devices.

**Acknowledgements:** We gratefully acknowledge Ronald Koder for the use of his facilities and equipment and Yun Yu and Michael Mirkin for AFM imaging of our samples. This work is supported by the National Science Foundation through Grants No. DMR-1410249 and DMR 1151783.

(1) Nichols, E. F.; Hull, G. F. *Am. Acad. Arts Sci.* **1903**, *38* (20), 559–599.

(2) A. Grinberg, A.; Luryi, S. *Phys. Rev. B* **1988**, *38* (1), 87–96.

(3) Gordon, J. P. *Phys. Rev. A* **1973**, *8* (1), 14–21.

(4) Goff, J. E.; Schaich, W. L. *Phys. Rev. B* **2000**, *61* (15), 10471–10477.

(5) S.D. Ganichev; Emel'yanov, S. A.; I.D. Yaroshetskii. *Sov. Phys. Semicond.* **1983**.

(6) Vengurlekar, A. S.; Ishihara, T. *Appl. Phys. Lett.* **2005**, *87* (9), 091118.

(7) Durach, M.; Rusina, A.; Stockman, M. *Phys. Rev. Lett.* **2009**, *103* (18), 186801.

(8) Noginova, N.; Yakim, A. V.; Soimo, J.; Gu, L.; Noginov, M. A. *Phys. Rev. B* **2011**, *84* (3), 035447.

(9) Noginova, N.; Rono, V.; Bezares, F. J.; Caldwell, J. D. *New J. Phys.* **2013**, *15* (11), 113061.

(10) Sheldon, M. T.; Groep, J. van de; Brown, A. M.; Polman, A.; Atwater, H. A. *Sci. Express* **2014**, *10* (1126).

(11) Hatano, T.; Ishihara, T.; Tikhodeev, S.; Gippius, N. *Phys. Rev. Lett.* **2009**, *103* (10), 103906.

(12) Akbari, M.; Onoda, M.; Ishihara, T. *Opt. Express* **2015**, *23* (2), 823.

(13) Kurosawa, H.; Ishihara, T. *Opt. Express* **2012**, *20* (2), 1561–1574.

(14) Knight, M. W.; Sobhani, H.; Nordlander, P.; Halas, N. J. *Science* **2011**, *332*, 702–704.

(15) Wang, F.; Melosh, N. A. *Nano Lett.* **2011**, *11*, 5426–5430.

(16) Bai, Q. *Opt. Express* **2015**, *23* (4), 5348.

(17) Karch, J.; Olbrich, P.; Schmalzbauer, M.; Zoth, C.; Brinsteiner, C.; Fehrenbacher, M.; Wurstbauer, U.; Glazov, M. M.; Tarasenko, S. A.; Ivchenko, E. L.; Weiss, D.; Eroms, J.; Yakimova, R.; Lara-Avila, S.; Kubatkin, S.; Ganichev, S. D. *Phys. Rev. Lett.* **2010**, *105* (22), 227402.

(18) Bliokh, K. Y.; Bliokh, Y. P. *Phys. Rev. Lett.* **2006**, *96* (7), 073903.

(19) Van Mechelen, T.; Jacob, Z. *arXiv.org* **2015**, 1–10.

(20) Bliokh, K. Y.; Smirnova, D.; Nori, F. *Science* **2015**, *348* (6242).

(21) Moocarme, M.; Kusin, B.; Vuong, L. T. *Opt. Mater. Express* **2014**, *4* (11), 645–648.

(22) Petersen, J.; Volz, J.; Rauschenbeutel, A. *Science* **2014**, *346* (6205), 67.

(23) Aiello, A.; Lindlein, N.; Marquardt, C.; Leuchs, G. *Phys. Rev. Lett.* **2009**, *103* (10), 100401.

(24) Dooghin, A. V.; Kundikova, N. D.; Liberman, V. S.; Zeldovich, B. Y. *Phys. Rev. A* **1992**, *45* (11), 8204–8208.

(25) Hermosa, N.; Nugrowati, A. M.; Aiello, A.; Woerdman, J. P. *Opt. Lett.* **2011**, *36* (16), 3200–3202.

(26) O'Neil, A.; MacVicar, I.; Allen, L.; Padgett, M. J. *Phys. Rev. Lett.* **2002**, *88* (5), 053601.

(27) Moocarme, M.; Dominguez-Juarez, J. L.; Vuong, L. T. *Nano Lett.* **2014**, *14*, 1178–1183.

(28) Bartlett, P. N.; Baumberg, J. J.; Coyle, S.; Abdelsalam, M. E. *Faraday Discuss.* **2004**, *125*, 117.

(29) Kim, M. H.; Im, S. H.; Park, O. O. *Adv. Funct. Mater.* **2005**, *15* (8), 1329–1335.

(30) Kelf, T.; Sugawara, Y.; Cole, R.; Baumberg, J.; Abdelsalam, M.; Cintra, S.; Mahajan, S.; Russell, A.; Bartlett, P. *Phys. Rev. B* **2006**, *74* (24), 245415.

(31) Vuong, L. T.; Adam, a. J. L.; Brok, J. M.; Planken, P. C. M.; Urbach, H. P. *Phys. Rev. Lett.* **2010**, *104* (8), 083903.

(32) Fernandez-Corbaton, I.; Zambrana-Puyalto, X.; Molina-Terriza, G. *Phys. Rev. A* **2012**, *86* (4), 1–14.

(33) Gorodetski, Y.; Niv, A.; Kleiner, V.; Hasman, E. *Phys. Rev. Lett.* **2008**, *101* (4), 043903.